\documentclass[prb,twocolumn,aps,showpacs,fixfloats]{revtex4}
\usepackage{graphicx}
\usepackage{bm}
\usepackage{amsmath,amssymb}
\usepackage{subfigure}
\usepackage{float}
\usepackage{latexsym}
\usepackage{epstopdf}
\usepackage{color}
\usepackage{enumerate}
\usepackage{pdfpages}

\newcommand{\s}{\scriptscriptstyle}

\begin{document}

\title {Evolution of the inhomogeneously-broadened  spin noise spectrum with ac drive  }

\author{Z. Yue  and M. E. Raikh }

\affiliation{Department of Physics and
Astronomy, University of Utah, Salt Lake City, UT 84112, USA}

\begin{abstract}
In the presence of random hyperfine fields, the noise spectrum, $\langle \delta s_{\omega}^2\rangle$,
of a spin ensemble represents a narrow peak centered at $\omega =0$ and a broad ``wing"
reflecting the distribution of the hyperfine fields. In the presence of an ac drive, the dynamics of a
single spin acquires additional harmonics at frequencies determined by both,
the drive frequency and the local field.
These harmonics are reflected as additional peaks in the
noise spectrum.
%of a given spin.
%and, correspondingly, the noise spectrum acquires new features.
We study how the {\em ensemble-averaged} $\langle \delta s_{\omega}^2\rangle$ evolves with the
drive amplitude, $\omega_{\s dr}$ (in the frequency units). Our main finding is that
additional peaks in the spectrum, caused by the drive, remain sharp even when $\omega_{\s dr}$ is
much smaller than the
typical hyperfine field. The reason is that the drive affects only
the spins for which the local Larmour
frequency is close to the
drive frequency. The shape of the low-frequency ``Rabi"-peak in
$\langle \delta s_{\omega}^2\rangle$ is universal with both, the position
and the width, being of the order of
$\omega_{\s dr}$. When the drive amplitude exceeds the width of the
hyperfine field distribution,
the noise spectrum transforms into a set of sharp peaks
centered at harmonics of the drive frequency.
\end{abstract}

\pacs{85.75.-d,72.25.Rb, 78.47.-p}
\maketitle

\hspace{-4cm}

\section{Introduction}
Common experimental techniques for the study of spin dynamics in semiconductors
include the polarization of luminescence upon optical spin
orientation\cite{PolarizationOfLuminescence} and the time-resolved
Faraday rotation\cite{FaradayRotation}.
Recently,  a third technique, spin noise spectroscopy, had been applied
to bulk semiconductors\cite{FirstInSemiconductors,NoiseInGaAs,NoiseInZnO}
and various semiconductor structures\cite{QuantumWell,dot2010,Li1,Li2}, see the
reviews Ref.~\onlinecite{review2010} and Ref.~\onlinecite{review2014}
for comprehensive literature.
Within the spin-noise technique, the  dynamics of spins manifests
itself via random modulation of  the
refraction indices for the left- and right-polarized light.
This modulation results in
%leads to a
random rotation angle of the
plane of polarization of the transmitted light.
The power Fourier  spectrum of these random
rotations is proportional
to the spectrum, $\delta s_\omega^2$, of the spin fluctuations.
Originally, the spin-noise measurements were
conducted on atomic vapors\cite{Classical,2004}.
With regard to  spin-noise, the principal difference
between the vapors and semiconductors is that all
spin-related frequencies  in  vapor are the same, while,
in semiconductors, these frequencies are strongly different
for different electrons. This is because, without external magnetic field,
each electron spin precesses around its individual hyperfine field
%electron spins are involved in precession even without external field.
%This is because the spins in semiconductor They
%precess around the random magnetic fields
created by  nuclei which are located within the extent of the
electron wave function\cite{Merkulov,Khaetskii}.
Importantly, the observation of  spin noise for
localized electrons and holes, see e.g. Refs.
%\onlinecite{NoiseInZnO},~\onlinecite{QuantumWell},~\onlinecite{dot2010}-\onlinecite{Li2}
\onlinecite{NoiseInZnO},~\onlinecite{dot2010}-\onlinecite{Li2},
%,~\onlinecite{Li1},~
 was possible
%in semiconductors has been reported\cite{FirstInSemiconductors}
even despite the strong inhomogeneous broadening.
When the applied magnetic field is much smaller than the
typical hyperfine field, the spin noise spectrum reflects the
distribution of the hyperfine fields\cite{Ivchenko,Glazov2014,Localized,Localized1}.
More precisely, the spectrum has the form\cite{Ivchenko}

\begin{multline}
\label{fromIvchenko}
\langle \delta s_{\omega}^2\rangle =
\\ \frac{\pi}{6}\left\{\Delta(\omega)\hspace{-1mm}+\hspace{-1mm}\int\limits_0^\infty
\hspace{-1mm}d\Omega_{\s N} F(\Omega_{\s N}) \Bigl[\Delta(\omega-\Omega_{\s N})+
\Delta(\omega+\Omega_{\s N})\Bigr] \right\},
\end{multline}
where
\begin{equation}
\Delta(\omega)=\frac{\tau_s}{\pi(1+\omega^2\tau_s^2)}
\end{equation}
is a Lorentzian and $\tau_s$ is the electron spin-relaxation time.
%The
Second term in Eq. (\ref{fromIvchenko}) represents the average
over the hyperfine fields, $\Omega_{\s N}$, distributed as\cite{Schulten}
\begin{equation}
\label{distribution}
F(\Omega_{\s N})=
\frac{4}{\sqrt{\pi}\delta_e^3}\Omega_{\s N}^2\exp \Bigl[-\frac{\Omega_{\s N}^2}{\delta_e^2}\Bigr].
\end{equation}
When the width, $\delta_e$, of the distribution exceeds $\tau_s^{-1}$,
the second term becomes $\frac{\pi}{6}\left[F(\omega)+F(-\omega)\right]$,
i.e. the shape of the noise spectrum reproduces the distribution of $\Omega_{\s N}$.
First term in Eq. (\ref{fromIvchenko}) represents a peak
centered at $\omega=0$. It originates from the fact that the spin
component parallel to the hyperfine field does not precess.

Very recently\cite{MostRelevant}, in the spin-noise experiment on a
vapor of $^{41} K$ alkali atoms, it was found
that the ac drive splits the noise spectrum into a
Mollow triplet. This splitting can be interpreted as a result
of modified spin dynamics in the
presence of  drive. In fact, such evolution of the noise spectrum
is in accord with the theoretical study of
Ref. \onlinecite{Galperin} where the noise of a driven two-level system  was considered.
It was demonstrated\cite{Galperin} that the drive-induced
additional harmonics in the dynamics of the two-level system
manifest themselves as additional peaks in the noise spectrum.

With regard to semiconductors, there is a question: what happens
to the spin noise spectrum in the presence of  drive when the
local hyperfine fields are widely
distributed? Since the positions of
the drive-related peaks depend on the local value of $\Omega_{\s N}$,
it might be expected that they average out.
Below we demonstrate that this is not the case.
It appears that the drive-related peaks
remain sharp after averaging.
The reason is that the major contribution to the averaged peaks comes
from the realizations of the
hyperfine field for which $\Omega_{\s N}$ is close to the drive frequency,
$\Omega_{\s dr}$.
More precisely, the domain of
$\Omega_{\s N}$ contributing to $\langle \delta s_{\omega}^2\rangle$ is
$|\Omega_{\s N}-\Omega_{\s dr}|\sim \omega_{\s dr}$, where $\omega_{\s dr}$ is the
drive amplitude.

If the frequency $\Omega_{\s dr}$ exceeds $\delta_e$ there are no
realizations of the hyperfine field
in resonance with drive.
In this case the drive has a strong effect on the noise
spectrum when the amplitude
$\omega_{\s dr}$ becomes comparable to $\Omega_{\s dr}$.
We will see that $\langle \delta s_{\omega}^2\rangle$ transforms
into a sequence of peaks at $\omega_n=n\Omega_{\s dr}$. The magnitudes of
the peaks behave essentially as $J_n^2(\omega_{\s dr}/\Omega_{\s dr})$,
where $J_n(x)$ is the Bessel function of the order $n$.

\begin{figure}
\includegraphics[width=80mm]{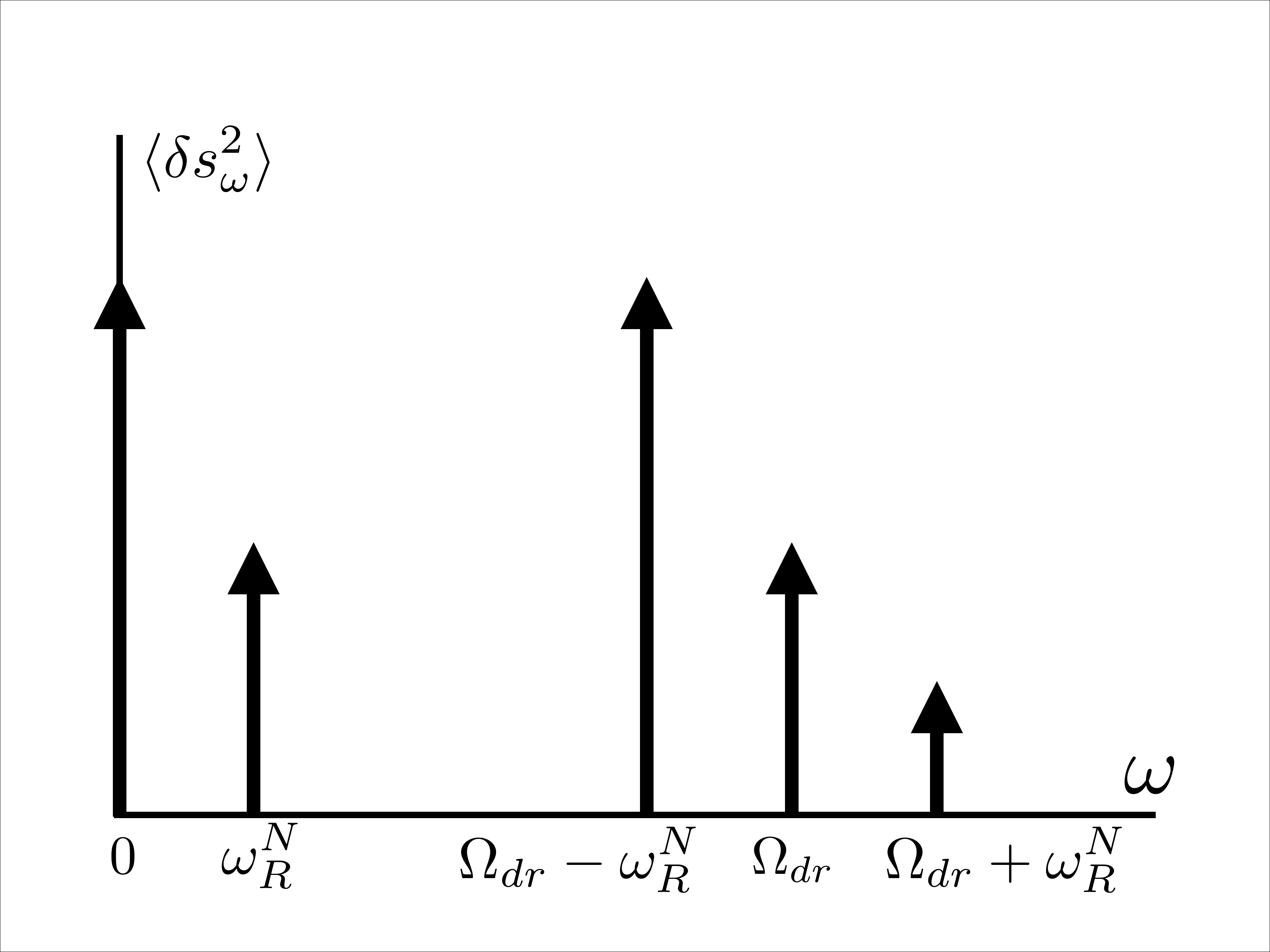}
\caption{ A cartoon of the noise spectrum for a
typical realization of the
hyperfine field in the presence of the ac drive. Due to drive, a zero-frequency peak develops a satellite
at $\omega =\omega_{\s R}^{\s N}$, Eq.~(\ref{Rabi}). The peak which, in the absence of drive, was located
at $\omega=\Omega_{\s N}$ shifts to $\omega= \Omega_{\s dr}-\omega_{\s R}^{\s N}$ and develops two satellites
at driving frequency and at $\omega=\Omega_{\s dr}+\omega_{\s R}^{\s N}$. The magnitudes of the satellites
scale with the drive amplitude as $\omega_{\s dr}^2$ and  $\omega_{\s dr}^4$, respectively. Both satellites are located to
the right from the main peak, which corresponds
to driving frequency exceeding $\Omega_{\s N}$.  }
\label{Fig1}
\end{figure}

\begin{figure}
\includegraphics[width=80mm]{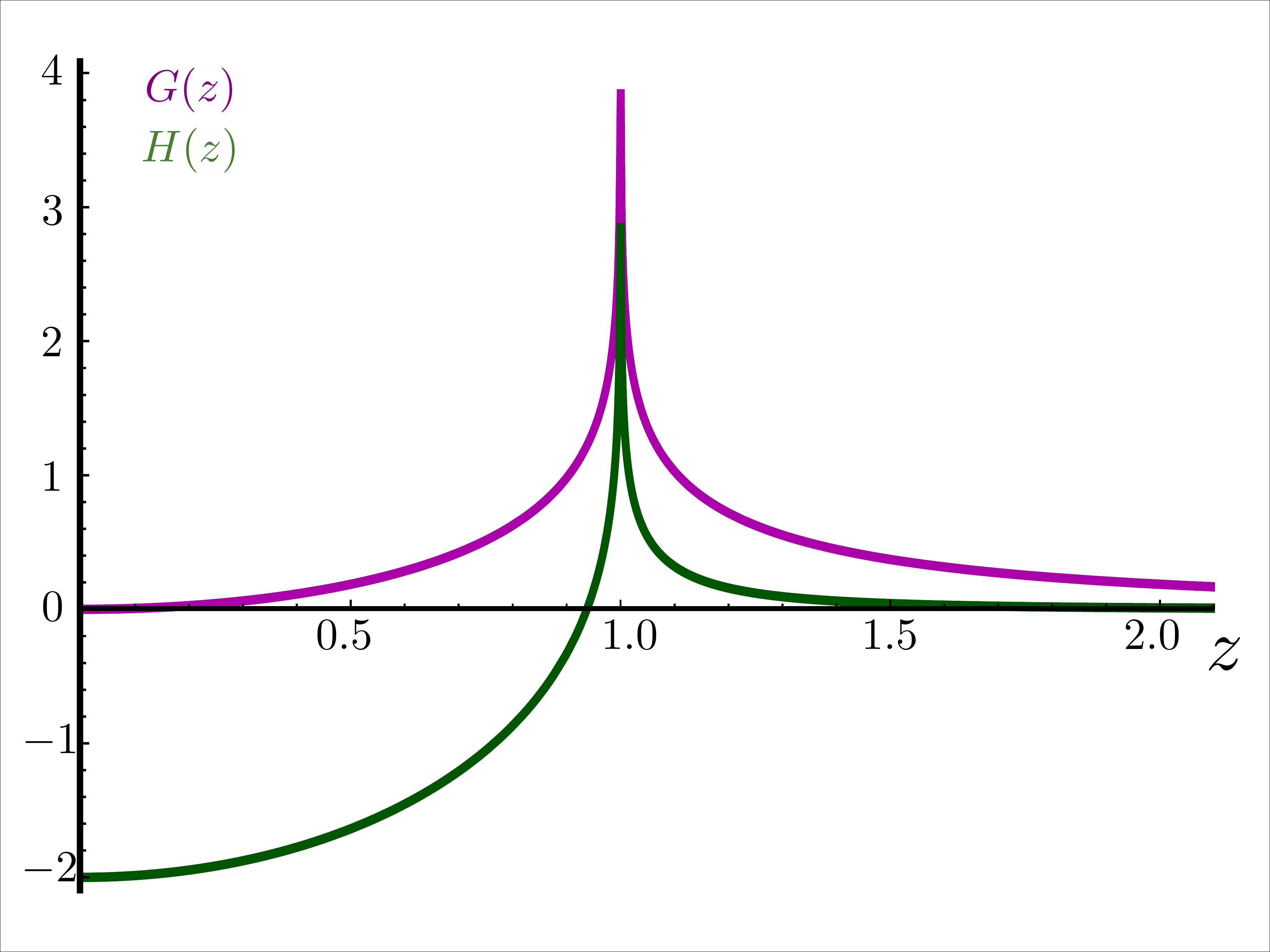}
\caption{(Color online) Dimensionless functions $G(z)$ (purple) and $H(z)$ (green) describing the shapes of the
peaks at Rabi frequency and at driving frequency in the averaged noise spectrum are plotted from
Eqs. (\ref{angular2}) and (\ref{angular4}), respectively.}
\label{Fig3}
\end{figure}

\section{General Expression for the noise spectrum with drive}
In the absence of spin relaxation and drive, the spin dynamics
is governed by the
equation  $\frac{d{\bf S}}{dt}={\bf \Omega}_{\s N}\times {\bf S}$,
which yields three modes:
\begin{eqnarray}
\label{matrix}
\begin{pmatrix}
 S_+\\S_z\\S_- \end{pmatrix}=A_+\begin{pmatrix} e^{i\Omega_{\s N} t} \\ 0 \\ 0 \end{pmatrix}+
 A_z\begin{pmatrix} 0 \\ 1 \\ 0\end{pmatrix}+A_-\begin{pmatrix} 0 \\ 0 \\ e^{-i\Omega_{\s N} t}
\end{pmatrix},
\end{eqnarray}
where $S_{\pm}=\frac{1}{\sqrt{2}}\left(S_x\pm iS_y\right)$. Here we assumed that the field ${\bm \Omega}_{\s N}$ is directed along the $z$-axis.
Spin relaxation is incorporated into the right-hand side of the equation of motion via a term, $-\frac{{\bf S}}{\tau_s}$.
As a result, the constants $A_{\pm}$ and $A_z$ in Eq. (\ref{matrix})
evolve with time simply as $\frac{dA_{\pm}}{dt}+ \frac{A_{\pm}}{\tau_s}=0$ and
$\frac{dA_z}{dt}+ \frac{A_z}{\tau_s}=0$. The fact that   the coefficients $A_{\pm}$ and $A_z$  decay as a simple exponent is
sufficient\cite{Ivchenko,Glazov2014,Localized,Localized1} to
present the noise spectrum as
\begin{align}
\label{nodrive}
\delta s_{z\omega}^2&=\frac{\pi}{2}\Delta(\omega)\\
\delta s_{x\omega}^2&=\delta s_{y\omega}^2=\frac{\pi}{4}\Bigl[\Delta(\omega-\Omega_{\s N})+
\Delta(\omega+\Omega_{\s N})\Bigr].
\end{align}

%\begin{align}
%S_{x'} &= S_x, \\
%S_{y'} &= S_y \cos \theta(t) + S_z \sin \theta(t) , \\
%S_{z'} &= -S_y \sin \theta(t) + S_z \cos \theta(t),
%\end{align}

In the presence of the ac drive the equation describing the spin dynamics assumes the form
\begin{equation}
\label{DynamicsWithDrive}
\frac{d{\bf S}}{dt}=\Bigl[{\bf \Omega}_{\s N}+2{\bm \omega}_{\s dr}\cos~(\Omega_{\s dr}t)\Bigr]\times {\bf S},
\end{equation}
where $2{\bm \omega}_{\s dr}$ and $\Omega_{\s dr}$ are the drive amplitude and frequency, respectively.
We will still assume that the hyperfine field is directed along the
$z$-axis. Then the component of drive, responsible for the spin  precession,
is $2\omega_{\s \perp dr}$, which is the projection of the driving field on the $x$-$y$ plane.
If the drive amplitude is much smaller than $\Omega_{\s N}$, the rotating wave approximation applies.
Then the solution of Eq.  (\ref{DynamicsWithDrive}) is well known since the classical paper
Ref. \onlinecite{Rabi}. We will cast this solution in the form, which, in the limit
$\omega_{\s dr} \rightarrow 0$, reduces to Eq. (\ref{matrix}). Namely:
%\begin{multline}
%\label{eigenvectors}
% \begin{pmatrix} S_+\\ S_z\\ S_- \end{pmatrix}= A_+\begin{pmatrix} \alpha_+ e^{i(\Omega_D-\sqrt{\omega_{\s \perp D}^2+(\Omega_D-\Omega_N)^2}) t} \\ \alpha_z e^{-i(\sqrt{\omega_{\s \perp D}^2+(\Omega_D-\Omega_N)^2}) t} \\ \alpha_- e^{-i(\Omega_D+\sqrt{\omega_{\s \perp D}^2+(\Omega_D-\Omega_N)^2}) t}\end{pmatrix}+ \\ A_z\begin{pmatrix} \beta_+ e^{i\Omega_D t} \\ \beta_z  \\ \beta_- e^{-i\Omega_D t} \end{pmatrix}+ A_-\begin{pmatrix} \gamma_+ e^{i(\Omega_D+\sqrt{\omega_{\s \perp D}^2+(\Omega_D-\Omega_N)^2}) t} \\ \gamma_z e^{i(\sqrt{\omega_{\s \perp D}^2+(\Omega_D-\Omega_N)^2}) t} \\ \gamma_- e^{-i(\Omega_D-\sqrt{\omega_{\s \perp D}^2+(\Omega_D-\Omega_N)^2}) t}\end{pmatrix},
%\end{multline}

\begin{multline}
\label{eigenvectors}
\begin{pmatrix} S_+\\ S_z\\ S_- \end{pmatrix} = A_+\begin{pmatrix} \alpha_+ e^{i(\Omega_{\s dr}-\omega_{\s R}^{\s N}) t}
\\ \alpha_z e^{-i\omega_{\s R}^{\s N} t} \\ \alpha_- e^{-i(\Omega_{\s dr}+\omega_{\s R}^{\s N}) t}\end{pmatrix}+
A_z\begin{pmatrix} \beta_+ e^{i\Omega_{\s dr} t} \\ \beta_z  \\ \beta_- e^{-i\Omega_{\s dr} t} \end{pmatrix}\\
 + A_-\begin{pmatrix} \gamma_+ e^{i(\Omega_{\s dr}+\omega_{\s R}^{\s N}) t} \\ \gamma_z e^{i\omega_{\s R}^{\s N} t} \\ \gamma_- e^{-i(\Omega_{\s dr} -\omega_{\s R}^{\s N}) t}\end{pmatrix},\quad\quad\quad\quad\quad
\end{multline}
where $\omega_{\s R}^{\s N}$ is the frequency of the Rabi oscillations defined as
\begin{equation}
\label{Rabi}
\omega_{\s R}^{\s N}=\left[\omega_{\s \perp dr}^2+(\Omega_{\s dr}-\Omega_{\s N})^2\right]^{1/2}.
\end{equation}
The relation between the coefficients for each mode of precession, say, between $\alpha_{+}$,
$\alpha_z$, and $\alpha_{-}$, follows from Eq. (\ref{DynamicsWithDrive})
%\begin{multline}
%\alpha_+=-\sqrt{\frac{\omega_{\perp D}^2}{2(\omega_{\perp D}^2+(\Omega_D-\Omega_N)^2)}}\\ \times \frac{\omega_{\perp D}}
% {\sqrt{2}(\Omega_D-\Omega_N-\sqrt{(\omega_{\perp D}^2+(\Omega_D-\Omega_N)^2)}},\\
%\alpha_z=\sqrt{\frac{\omega_{\perp D}^2}{2(\omega_{\perp D}^2+(\Omega_D-\Omega_N)^2)}}, \\
%\alpha_-=-\sqrt{\frac{\omega_{\perp D}^2}{2(\omega_{\perp D}^2+(\Omega_D-\Omega_N)^2)}}\\ \times \frac{\omega_{\perp D}}{\sqrt{2}(\Omega_D-\Omega_N+\sqrt{(\omega_{\perp D}^2+(\Omega_D-\Omega_N)^2)}}.
%\end{multline}
\begin{eqnarray}
\label{alpha}
\alpha_+=-\frac{\omega_{\s \perp dr}^2}{2\omega_{\s R}^{\s N} (\Omega_{\s dr}-\Omega_{\s N}-\omega_{\s R}^{\s N})}\\
%\alpha_+=\frac{\Omega_D-\Omega_N+\omega_R^N}{2\omega_R^N}\\
\alpha_z=\frac{\omega_{\s \perp dr}}{\sqrt{2}\omega_{\s R}^{\s N}}\\
\alpha_-=-\frac{\omega_{\s \perp dr}^2}{2\omega_{\s R}^{\s N} (\Omega_{\s dr}-\Omega_{\s N}+\omega_{\s R}^{\s N})}.
\end{eqnarray}
The magnitudes of $\alpha_{+}$,
$\alpha_z$, $\alpha_{-}$ are chosen in such a way that the corresponding eigenvector
in Eq. (\ref{eigenvectors}) is normalized. It is easy to see that,
as the drive decreases, $\alpha_+$ approaches one, while $\alpha_z$ and $\alpha_-$ vanish.
This applies for
 $\Omega_{\s dr} > \Omega_{\s N}$. For the opposite relation, $\alpha_+$
 vanishes upon decreasing drive, while  $\alpha_-$ approaches one.
 In a similar way, for remaining two eigenvectors we have
\begin{align}
\label{relations}
 \beta_+=\beta_-=\alpha_z\\
\beta_z=\frac{\Omega_{\s dr}-\Omega_{\s N}}{\omega_{\s R}^{\s N}}\\
\gamma_+=\alpha_-,~~\gamma_-=\alpha_+,~~\gamma_z=\alpha_z.
\end{align}
In the presence of spin relaxation the coefficients
$A_{\pm}$ and $A_z$ in Eq. (\ref{eigenvectors})
satisfy the same equation as in the absence of drive.
This allows to establish the form of the noise spectrum,
$\delta s_\omega^2=\frac{1}{3}(\delta s_{\omega z}^2+\delta s_{\omega +}^2+\delta s_{\omega - }^2)$,
of the driven system in the
same way as Eq. (\ref{nodrive})  followed from Eq. (\ref{matrix}). One has
\begin{eqnarray}
\label{spectrum}
\delta s_\omega^2=\frac{\pi}{6}\Bigl\{ \beta_z^2 \Delta(\omega)+ \gamma_z^2 \Delta(\omega-\omega_{\s R}^{\s N})+\alpha_z^2
\Delta(\omega+\omega_{\s R}^{\s N})\nonumber \\  +\beta_+^2 \Delta(\omega-\Omega_{\s dr})+\beta_-^2 \Delta(\omega+\Omega_{\s dr})+
\nonumber \\ \gamma_+^2 \Delta(\omega-\Omega_{\s dr}-\omega_{\s R}^{\s N}) +\alpha_-^2 \Delta(\omega+\Omega_{\s dr}+\omega_{\s R}^{\s N}) +
\nonumber \\ \alpha_+^2 \Delta(\omega-\Omega_{\s dr}+\omega_{\s R}^{\s N})+ \gamma_-^2 \Delta(\omega+\Omega_{\s dr}-\omega_{\s R}^{\s N})\Bigr\}.
\end{eqnarray}
It is a direct consequence of normalization of the eigenvectors in Eq. (\ref{eigenvectors})
that the area $\int d\omega \delta s_{\omega}^2$ does not depend on the drive. Four  groups
of terms corresponding to the four lines in Eq. (\ref{spectrum}) can be interpreted as
follows. The low-frequency peak $\propto \Delta(\omega)$  in the presence of drive
develops two satellites at $\omega=\pm \omega_{\s dr}$. From the relation $\beta_z^2+\gamma_z^2+\alpha_z^2=1$, which can be easily checked using Eqs. (\ref{alpha}), (\ref{relations}), it follows that the noise power gets redistributed between the three peaks.
The peak which, in the absence of drive, was located at $\omega =\Omega_{\s N}$
shifts to the position $\omega=\Omega_{\s dr}-\omega^{\s R}_{\s N}$.
%Correspondingly, the $\omega =\Omega_N$ peak  in the absence of drive shifts
%to the position $\omega=\Omega_D-\omega^R_N$
It also follows from Eq. (\ref{spectrum}) that this peak  develops two satellites
%at driving
at higher frequencies
$\omega=\Omega_{\s dr}$ and at $\omega=\Omega_{\s dr}+\omega^{\s R}_{\s N}$ with magnitudes $\beta_+^2$ and $\gamma_+^2$, respectively. Again, the net noise power in these three peaks does not depend on drive.

Suppose that the drive is weak,  $\omega_{\s \perp dr} \ll \Omega_{\s N}$. For a typical  realization of the hyperfine field the difference $\Omega_{\s N} -\Omega_{\s dr}$ is much bigger than $\omega_{\s \perp dr}$. Then the relative magnitude of the satellites of the zero-frequency peak is
equal to $\omega_{\s \perp dr}^2/2(\Omega_{\s N} -\Omega_{\s dr})^2$. With regard to the peak at
$\omega=\Omega_{\s N}$, its shift due to drive is small, namely, $\omega_{\s \perp dr}^2/2(\Omega_{\s N} -\Omega_{\s dr})$.
% quadratic in $\omega_{\s \perp dr}$.
The magnitudes  of these satellites evolve differently with drive:
while the  satellite at $\omega=\Omega_{\s dr}$ grows as
$\omega_{\s \perp dr}^2/2(\Omega_{\s N} -\Omega_{\s dr})^2$, the satellite at
$\omega=\Omega_{\s D}+\omega^{\s R}_{\s N} \approx 2\Omega_{\s dr}-\Omega_{\s N}$
has a much smaller relative magnitude $\sim \omega_{\s \perp dr}^4/16(\Omega_{\s N} -\Omega_{\s dr})^4$.
A generic noise spectrum is illustrated in Fig. \ref{Fig1}.
Overall,   the effect of drive on the  noise spectrum for a {\em typical} $\Omega_{\s N}$ is weak.
In addition, the positions of all satellites, except the peak at $\omega=\Omega_{\s dr}$,
depend on $\Omega_{\s N}$, i.e. these positions  are random. It is not clear whether these
satellites manifest themselves in the {\em ensemble-averaged} noise spectrum.
As we will see in the next Section, the averaging preserves the drive-induced
peaks in the noise spectrum.
The reason is that the realizations of $\Omega_{\s N}$, which survive the averaging, are the
those in ``resonance" with drive. For such realizations, with
$|\Omega_{\s N} -\Omega_{\s dr}|\sim \omega_{\s \perp dr}$, the magnitudes of the
satellites are anomalously big. This compensates for small statistical weight of the resonant configurations.

\begin{figure}
\includegraphics[width=80mm]{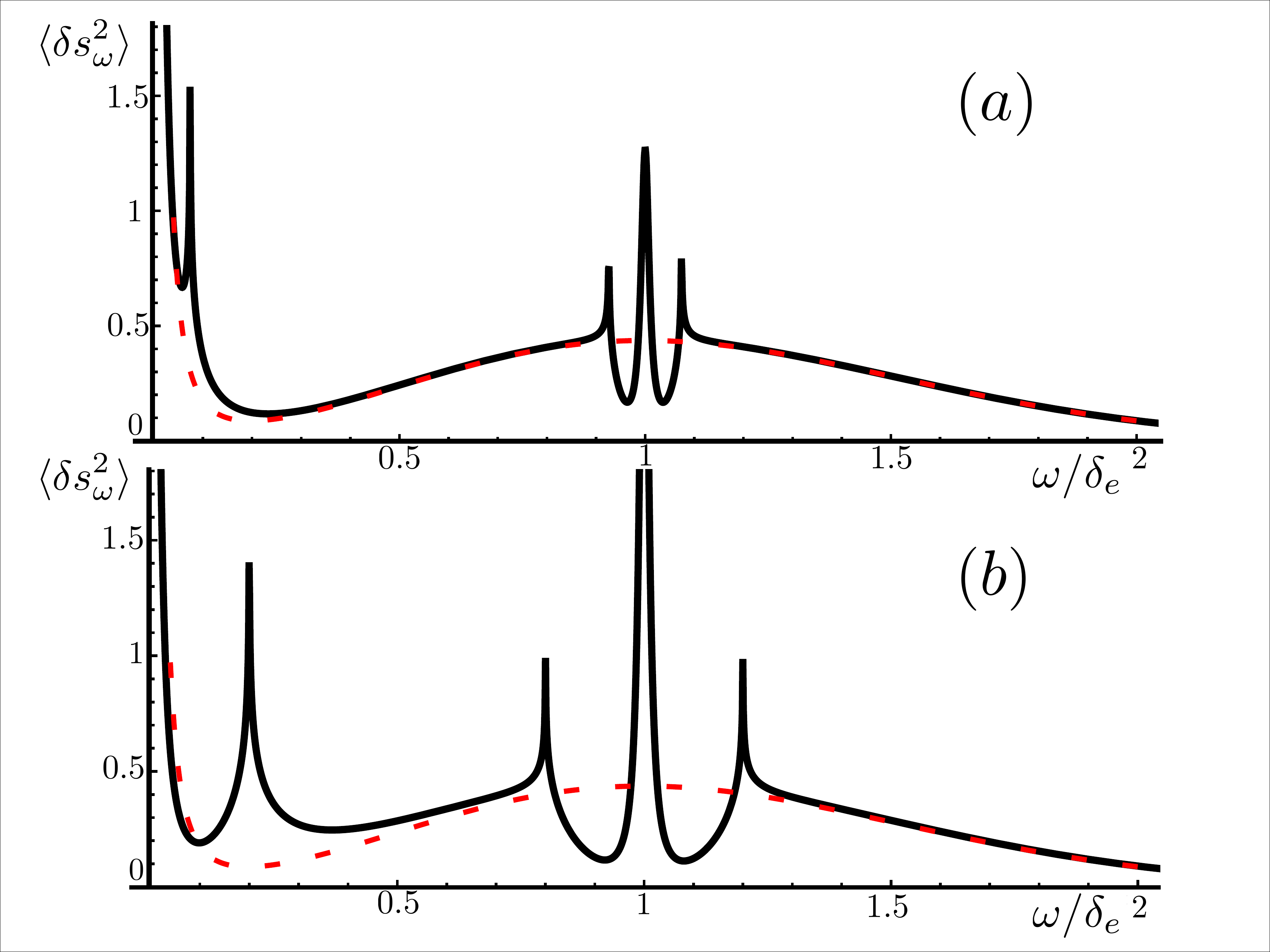}
\caption{(Color online) Averaged noise spectra are plotted from Eq. (\ref{final}) for driving frequency,
$\Omega_{\s dr}$, corresponding to the maximum of the hyperfine field distribution and
two amplitudes of the drive: $2\omega_{\s dr}=0.15\delta_e$ (a) and $2\omega_{\s dr}=0.4\delta_e$ (b).
Even for weak drive the drive-induced satellites in the averaged spectra are  well-pronounced.
The width of the central peak at $\omega=\Omega_{\s dr}$ is $\tau_s^{-1}=10^{-2}\delta_e$.
The spectra in the absence of drive, illustrating the conservation of the net noise power, are shown with red lines.}
\label{Fig2}
\end{figure}

\section{Ensemble averaging}
We will perform the averaging over hyperfine fields in two steps. Firstly, we will  average over the magnitudes, $\Omega_{\s N}$,
with distribution function $F(\Omega_{\s N})$. and then the
As a second step, we will average over directions of $\bm{\Omega}_{\s N}$
with respect to the driving field, $\bm{\omega}_{\s dr}$.
\subsection{Averaging over the magnitudes of hyperfine fields}
Averaging of the first term in Eq. (\ref{spectrum}) is straightforward, since  $\Delta(\omega)$ does not depend on $\Omega_{\s N}$. Using the definition Eq. (\ref{relations}),
we have
\begin{equation}
\label{average1}
\frac{\pi}{6}\langle\beta_z^2\rangle \Delta(\omega)=\frac{\pi}{6}\Bigl[1-\int\limits_0^{\infty}
d\Omega_{\s N}\frac{\omega_{\s \perp dr}^2F(\Omega_{\s N})}{(\Omega_{\s N}-\Omega_{\s dr})^2+\omega_{\s \perp dr}^2} \Bigr]\Delta(\omega).
\end{equation}
Taking into account that the typical difference $(\Omega_{\s N}-\Omega_{\s dr})\sim \delta_e$
is much bigger than the drive amplitude, $\omega_{\perp dr}$, we get
\begin{equation}
\label{average11}
\frac{\pi}{6}\Bigl[1-\pi\omega_{\s \perp dr}F(\Omega_{\s dr})\Bigr]\Delta(\omega).
\end{equation}
Thus, the reduction of the magnitude of the zero-frequency peak due to drive is {\em linear}
in drive amplitude and comes from the ``resonant" realizations of the hyperfine fields for
which $\Omega_{\s N}\approx \Omega_{\s dr}$.

It is less trivial to realize that the peaks at $\omega=\pm \omega_{\s R}^{\s N}$ described by the
second and third terms in Eq. (\ref{spectrum})  remain sharp upon averaging even though
their positions depend on $\Omega_{\s N}$.
The expression for the average of  the second term reads
\begin{equation}
\label{average2}
\frac{\pi}{6}\langle\gamma_z^2 \Delta(\omega-\omega_{\s R}^{\s N})\rangle=
\frac{\pi}{12}\omega_{\s \perp dr}^2
\int\limits_0^{\infty} d\Omega_{\s N} \frac{\Delta(\omega-\omega_{\s R}^{\s N})
F(\Omega_{\s N})}{\left(\omega_{\s R}^{\s N}\right)^2}.
\end{equation}
At this point we make use of the fact that the width of the Lorentzian,
$\Delta(\omega-\omega_{\s R}^{\s N})$, is much smaller than the width of the distribution
function. Firstly, this allows to set
$\omega_{\s R}^{\s N}=\omega$ in the denominator.
Secondly, the values $\Omega_{\s N}$ that contribute to the integral are close to
\begin{equation}
\label{average22}
\Omega_{\s N}=\Omega_{\s dr}\pm \sqrt{\omega^2-\omega_{\s \perp dr}^2}.
\end{equation}
Upon switching to the integration over $\omega_{\s R}^{\s N}$ and
taking into account that
\begin{equation}
\label{average23}
\frac{d\omega_{\s R}^{\s N}}{d\Omega_{\s N}}=
\frac{\sqrt{\left(\omega_{\s R}^{\s N}\right)^2-\omega_{\s \perp dr}^2}}{\omega_{\s R}^{\s N}}=
\frac{\sqrt{\omega^2-\omega_{\s \perp dr}^2}}{\omega},
\end{equation}
we find
\begin{multline}
\label{average24}
\frac{\pi}{6}\langle\gamma_z^2 \Delta(\omega-\omega_{\s R}^{\s N})\rangle=\frac{\pi}{12}
\frac{\omega_{\s \perp dr}^2}{\omega\sqrt{\omega^2-\omega_{\s \perp dr}^2}}\\ \times \Bigl[F\Bigl(\Omega_{\s dr}-
\sqrt{\omega^2-\omega_{\s \perp dr}^2}\Bigr)+F\Bigl(\Omega_{\s dr}+\sqrt{\omega^2-\omega_{\s \perp dr}^2}\Bigr)\Bigr].
\end{multline}
Since the driving frequency is much bigger than the driving amplitude,
both arguments in
 the distribution function can be replaced by $\Omega_{\s N}$.
 With regard to the frequency dependence, Eq. (\ref{average24})
 exhibits an integrable divergence near $\omega \approx \omega_{\s \perp dr}$.
Most importantly, the averaged peak falls off rapidly with $\omega$ as the difference
$\omega-\omega_{\s \perp dr}$ increases. From the area conservation,
it follows from Eq. (\ref{average11}) that the area under the peak Eq. (\ref{average24})
should be equal to $\frac{\pi^2}{12}\omega_{\s \perp dr}F(\Omega_{\s dr})$.
On the other hand, from Eq. (\ref{average24}) we see that this area comes from the domain
$(\omega-\omega_{\s \perp dr}) \sim \omega_{\s \perp dr}\ll \delta_e$,
i.e. the area  conservation is ensured
{\em locally}.
This supports our statement that the averaged peak  remains narrow.

Two terms on the second line of Eq. (\ref{spectrum})
describe the peaks at $\omega=\pm \Omega_{\s dr}$.
Similar to $\Delta(\omega)$ peak, their shape is not affected by the ensemble averaging.
Averaging of the magnitude is completely analogous to that for $\Delta(\omega)$
peak since $\beta_z^2+2\alpha_z^2=1$. Thus, the contribution of these peaks to the noise
spectrum is given by
\begin{equation}
\label{average31}
\frac{\pi^2}{12}\omega_{\s \perp dr}F(\Omega_{\s dr})\Delta(\omega \pm \Omega_{\s dr}),
\end{equation}
and grows linearly with the drive amplitude.

The last two lines in Eq. (\ref{spectrum}) describe the peaks
in the noise spectrum at frequencies
$\omega=\pm \Omega_{\s dr}\pm \omega_{\s R}^{\s N}$. Firstly, we note that the magnitudes of
all four peaks are equal to each other. This follows from the fact that
these magnitudes, $\gamma_{\pm}^2$ and $\alpha_{\pm}^2$ are determined by the values of $\Omega_{\s N}$
for which the arguments of the corresponding Lorentzians are zero.
Now the equality of all peak magnitudes follows from the relation
\begin{equation}
\label{average41}
\left.\gamma_{+}\right|_{\Omega_{\s N}=\omega-\omega_{\s R}^{\s N}}=
\left.\alpha_{+}\right|_{\Omega_{\s N}=\omega+\omega_{\s R}^{\s N}},
\end{equation}
which is easy to check using Eqs. (\ref{alpha}) and (\ref{relations}).
Focusing on positive $\omega$,
the averaging over $\Omega_{\s N}$ is easy to perform by replacing
Lorentzians by corresponding
$\delta$-functions and using the following identities
\begin{multline}
\label{average42}
\delta\Bigl(\omega-\Omega_{\s dr}-\omega_{\s R}^{\s N}\Bigr)+\delta\Bigl(\omega-\Omega_{\s dr}+\omega_{\s R}^{\s N}\Bigr)
=\\ 2\omega_{\s R}^{\s N}\delta\Bigl((\omega-\Omega_{\s dr})^2-(\omega_{\s R}^{\s N})^2\Bigr)
=\frac{\omega_{\s R}^{\s N}}{\sqrt{(\omega-\Omega_{\s dr})^2-\omega_{\s \perp dr}^2}}\\
\times\Bigl[\delta\Bigl(\sqrt{(\omega-\Omega_{\s dr})^2-\omega_{\s \perp dr}^2}-(\Omega_{\s dr}-\Omega_{\s N})\Bigr)+
\\ \delta\Bigl(\sqrt{(\omega-\Omega_{\s dr})^2-\omega_{\s \perp dr}^2}+(\Omega_{\s dr}-\Omega_{\s N})\Bigr)\Bigr].
\end{multline}
Upon averaging, the last two $\delta$-functions pick the distribution $F(\Omega_{\s N})$
at the values $\Omega_{\s N}=\Omega_{\s dr}\pm \sqrt{(\omega-\Omega_{\s dr})^2+\omega_{\s \perp dr}^2}$.
The resulting average shape of the two peaks
at $\omega=\Omega_{\s dr}\pm \omega_{\s R}^{\s N}$ reads

\begin{multline}
\label{average43}
\frac{\pi\omega_{\s \perp dr}^4}{24|\omega-\Omega_{\s dr}|\sqrt{(\omega-\Omega_{\s dr})^2-
\omega_{\s \perp dr}^2}}\\ \times \Biggl[\frac{F\Bigl(\Omega_{\s dr}-\sqrt{(\omega-\Omega_{\s dr})^2-
\omega_{\s \perp dr}^2}\Bigr)}{(\omega-\Omega_{\s dr}+\sqrt{(\omega-\Omega_{\s dr})^2-\omega_{\s \perp dr}^2})^2}\\
+\frac{F\Bigl(\Omega_{\s dr}+\sqrt{(\omega-\Omega_{\s dr})^2-\omega_{\s \perp dr}^2}\Bigr)}{(\omega-\Omega_{\s dr}-
\sqrt{(\omega-\Omega_{\s dr})^2-\omega_{\s \perp dr}^2})^2}\Biggr].
\end{multline}
Note now, that for in the limit $\omega_{\perp dr}\rightarrow 0$ the above expression reproduces the
second term in Eq. (\ref{fromIvchenko}), i.e. the background noise spectrum in the absence of drive.
Formally this follows from the fact that either the first denominators in  Eq. (\ref{average43})
(for $\omega <\Omega_{\s dr}$) or the  second denominator (for $\omega <\Omega_{\s dr}$)
becomes small, $\propto \omega_{\s \perp dr}^4$.
In order to isolate the drive-related peaks from Eq. (\ref{average43})
one has to subtract
\begin{equation}
\label{average44}
\frac{\pi}{6}F\Bigl(\Omega_{\s dr}+\frac{\omega-\Omega_{\s dr}}{|\omega-
\Omega_{\s dr}|}\sqrt{(\omega-\Omega_{\s dr})^2-\omega_{\s \perp dr}^2}\Bigr)
\end{equation}
from Eq. (\ref{average43}). This  subtracted term is a smooth function of $\omega$.
On the other hand, after the subtraction, Eq. (\ref{average43}) would describe two narrow peaks
at $\omega+\Omega_{\s dr}\pm \omega_{\s \perp dr}$. This again allows to set
$\Omega_{\s N}=\Omega_{\s dr}$
in the argument
of distribution function. It is convenient to cast the final result in the form
\begin{equation}
\label{average45}
\frac{\pi}{12}\Bigl[\frac{|\omega-\Omega_{\s dr}|}{\sqrt{(\omega-\Omega_{\s dr})^2-
\omega_{\s \perp dr}^2}}+\frac{\sqrt{(\omega-\Omega_{\s dr})^2-\omega_{\s \perp dr}^2}}
{|\omega-\Omega_{\s dr}|}-2\Bigr] F(\Omega_{\s dr}).
\end{equation}
It is worth noting that the peaks described by Eq.  (\ref{average43}), having
the same width $\sim \omega_{\s \perp dr}$, are
``shaper" than the peak Eq. (\ref{average24}) at  the Rabi frequency.
They decay as $\omega_{\s \perp dr}^4/(\omega -\Omega_{\s dr})^4$, while the peak at
the Rabi frequency decays as $\omega_{\s \perp dr}^2/\omega^2$.
%The peaks described by Eqs. (\ref{average11}), (\ref{average24}),
%(\ref{average31}), and  (\ref{average43})       are illustrated in Fig. {\bf 2}.

\subsection{Averaging over orientations of  hyperfine fields}
The shape of the peaks in the noise spectrum derived above depends on $\omega_{\s \perp dr}$, the projection
of the driving field on the plane normal to the local hyperfine field. If the angle between ${\bm \Omega}_{\s N}$
and ${\bm \omega}_{\s dr}$ is $\theta$, then $\omega_{\s \perp dr}=\omega_{\s dr} \sin\theta$. To find the
ensemble-averaged shape of the noise spectrum with drive one has to average over $\theta$ all four contributions
Eqs. (\ref{average11}), (\ref{average24}),(\ref{average31}), and (\ref{average45})  as $\frac{1}{2}\int_0^\pi d\theta \sin\theta ~(......)$.

Averaging of Eqs. (\ref{average11}), (\ref{average31}) simply reduces to the replacement of
$\omega_{\s \perp dr}$ by $\frac{\pi}{4}\omega_{\s dr}$ without affecting the Lorentzian shapes of the narrow peaks.
The prime effect of averaging of Eqs. (\ref{average24}) and (\ref{average45}) is the rounding of
$1/\sqrt{\omega-\omega_{\s \perp dr}}$ and $1/\sqrt{\omega-\Omega_{\s dr}\pm \omega_{\s \perp dr}}$ anomalies.
These anomalies do not disappear completely but become logarithmical. Both averages
can be evaluated analytically. The averaging of Eq. (\ref{average24}) yields
\begin{equation}
\label{angular1}
\frac{\pi}{6}F(\Omega_{\s dr})~G\Bigl(\frac{\omega}{\omega_{\s dr}}\Bigr),
\end{equation}
where the dimensionless function $G(z)$ describing the averaged shape is given by
\begin{equation}
\label{angular2}
G(z)\hspace{-1mm}=\frac{1}{2}\hspace{-1mm}\int\limits_0^\pi \hspace{-1mm}d\theta~\frac{\sin^3\theta}{z\sqrt{z^2-\sin^2\theta}}
=\hspace{-1mm}\begin{cases}
 \frac{z^2+1}{2z}\ln\frac{z+1}{\sqrt{1-z^2}}-\frac{1}{2},~~ z<1, \\
 \frac{z^2+1}{4z}\ln\frac{z+1}{z-1}-\frac{1}{2},~~~~~ z>1.
 \end{cases}
\end{equation}
The divergence near $z=1$ should be cut off at $(1-z)\sim 1/\omega_{\s dr}\tau_s$. The large-$z$ behavior of
Eq. (\ref{angular2}) is $G(z)\approx 2/3z^2$.

The result of averaging of Eq. (\ref{average45}) can be presented in the form similar to Eq. (\ref{angular2})
\begin{equation}
\label{angular3}
\frac{\pi}{12}F(\Omega_{\s dr})H\Bigl(\frac{|\omega-\Omega_{\s dr}|}{\omega_{\s dr}}\Bigr),
%\Bigl[G\Bigl(\frac{|\omega-\Omega_D|}{\omega_D}\Bigr)-
%H\Bigl(\frac{|\omega-\Omega_D|}{\omega_D}\Bigr)-2\Bigr],
\end{equation}
where the function $H(z)$ is defined as
\begin{equation}
H(z)=\frac{1}{2}\int\limits_0^\pi d\theta~ \frac{2z\sin\theta -\sin^3\theta}{\sqrt{z^2-\sin^2\theta}}-2.
\end{equation}
Similarly  to Eq. (\ref{angular2}), the integral can be evaluated analytically yielding
\begin{equation}
\label{angular4}
H(z)= \begin{cases}
 \frac{3z^2-1}{2z}\ln\frac{z+1}{\sqrt{1-z^2}}-\frac{3}{2},~~~~ z<1, \\
 \frac{3z^2-1}{4z}\ln\frac{z+1}{z-1}-\frac{3}{2},~~~~~~~ z>1.
 \end{cases}
\end{equation}
The large-$z$ behavior of the combination in the square brackets is $\propto 1/z^4$.
Dimensionless functions $G(z)$ and $H(z)$ are plotted in Fig. \ref{Fig3}.

Combining all the above, the final result for the ensemble-averaged noise spectrum of the driven system
can be cast in the form
%\begin{multline}
\begin{widetext}
\begin{equation}
\label{final}
\langle \delta s_{\omega}^2\rangle = \frac{\pi}{6}\Bigl[1-\frac{\pi^2\omega_{\s dr}}{4} F(\Omega_{\s dr})\Bigr]\Delta(\omega)+\frac{\pi}{6}F(\omega)\\ +\frac{\pi}{6}F(\Omega_{\s dr})\Biggl\{G\Bigl(\pm \frac{\omega}{\omega_{\s dr}}\Bigr) +\frac{\pi^2}{8}\omega_{\s dr}\Delta(\omega\pm \Omega_{\s dr})+\frac{1}{2}H\Bigl(\frac{|\omega \pm \Omega_{\s dr}|}{\omega_{\s dr}}\Bigr)\Biggr\}.
\end{equation}
%\end{multline}
\end{widetext}
It is natural that the drive-related contributions are proportional to the density, $F(\Omega_{\s dr})$, of the  resonant realizations of hyperfine fields. The net effect of  drive on the noise spectrum is maximal if $\Omega_{\s dr}$ is chosen near the maximum of the distribution $F(\Omega_{\s N})$.
Then, at frequencies $\omega \sim \omega_{\s dr}$, the background noise is determined by $F(\omega)\propto \omega^2$ and is much {\em weaker}
than the low-frequency peak due to drive. The area under all three peaks in the second line of
Eq. (\ref{final}) is $\sim \omega_{\s dr}$. In this regard, the weakness of drive means that the  portion of
the noise spectrum affected by drive is relatively small. However, within this portion, the spectrum is fully dominated by
drive, since, for the resonant realizations, the drive changes the spin dynamics  completely.
Formally, it is the consequence of Eq. (\ref{Rabi}), that for $(\Omega_{\s N}-\Omega_{\s dr})\lesssim \omega_{\s dr}$
all the frequencies $\omega_{\s R}^{\s N}$ are close to $\omega_{\s dr}$. The evolution of the noise spectrum
with drive amplitude is illustrated in Fig. \ref{Fig2}.

\begin{figure}
\includegraphics[width=80mm]{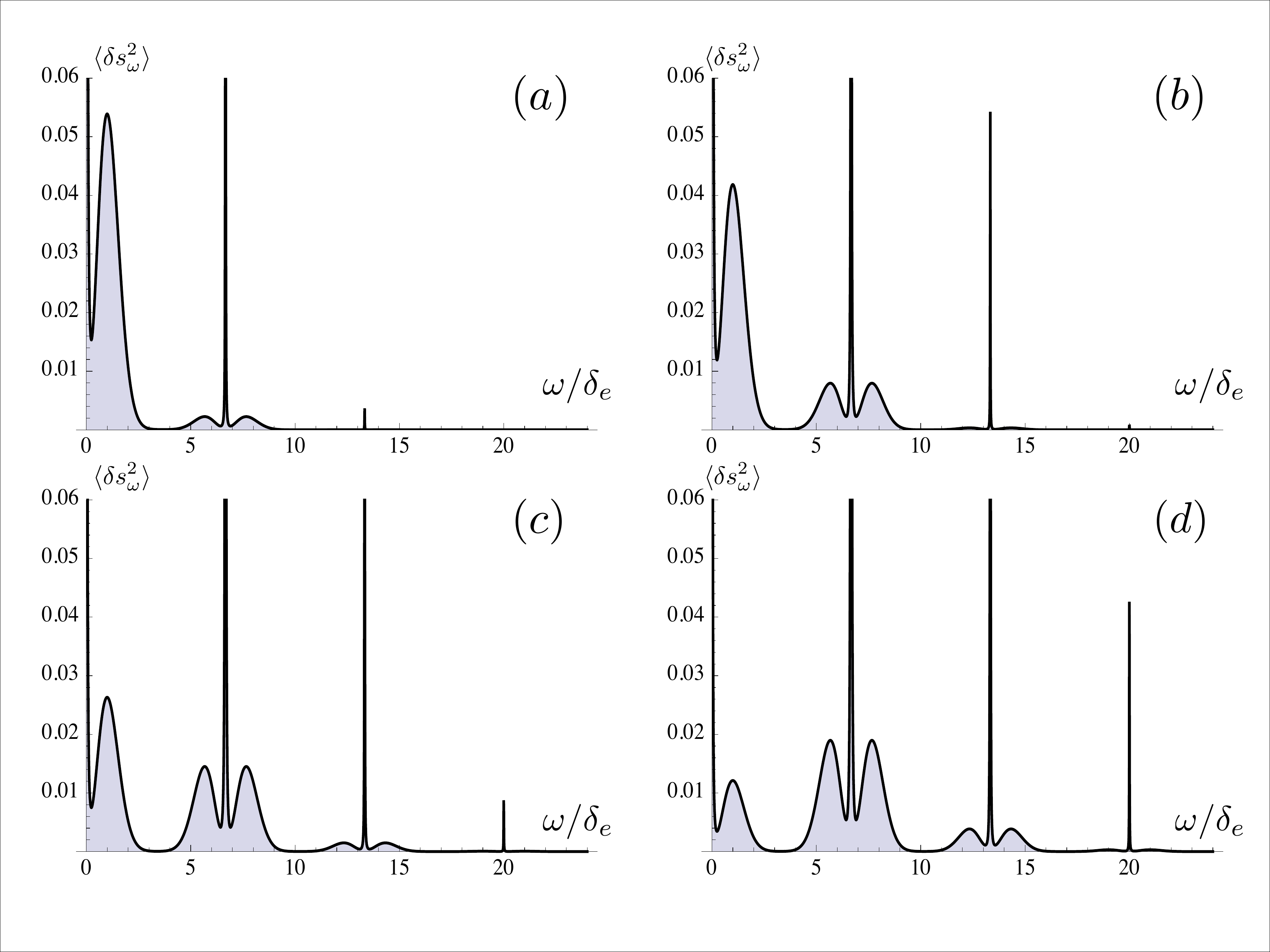}
\caption{(Color online) Evolution of the averaged noise spectra in the case of a fast
drive, $\Omega_{\s dr}=6.7\delta_e$, with drive amplitude, $\omega_{\s dr}$.
The values of the amplitudes are $\omega_{\s dr}=0.4\Omega_{\s dr}$(a), $\omega_{\s dr}=0.8\Omega_{\s dr}$ (b),
$\omega_{\s dr}=1.2\Omega_{\s dr}$ (c),
and $\omega_{\s dr}=1.6\Omega_{\s dr}$(d).
Narrow peaks have the width $\tau_s^{-1}=0.0067\delta_e$. Satellites of the zero-frequency peak at
$\omega=n\Omega_{\s dr}$ gradually develop upon increasing $\omega_{\s dr}$. Since the value
$J_0\Bigl(\frac{\omega_{\s dr}}{\Omega_{\s dr}}\Bigr)$ is close to $1$ for all $\omega_{\s dr}$ the low-frequency parts
 of the spectra has the same shape as in the absence of drive.    }
\label{4a}
\end{figure}
\begin{figure}
\includegraphics[width=80mm]{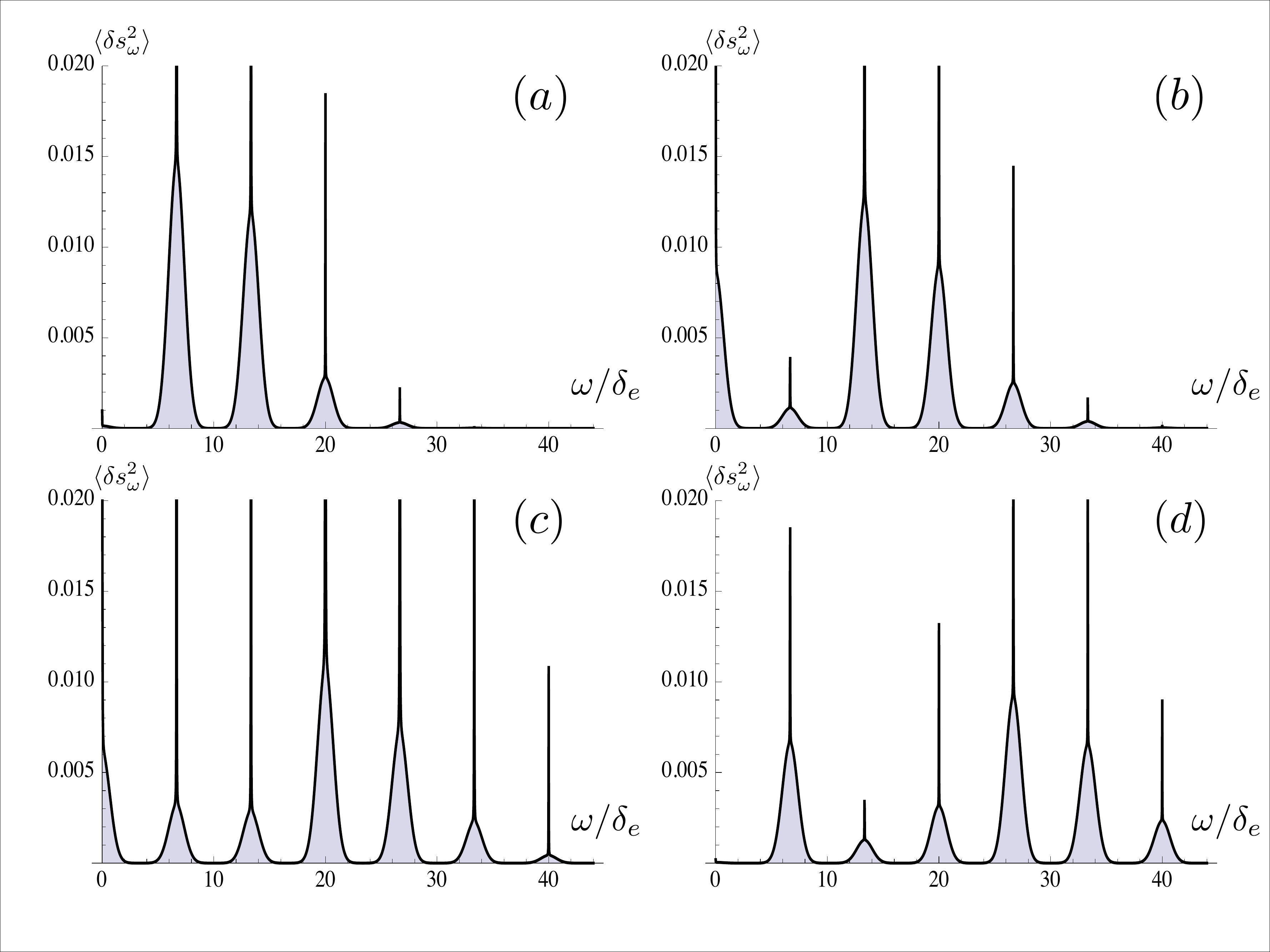}
\caption{(Color online)
Same as Fig. \ref{4a} for stronger drive amplitudes $\omega_{\s dr}=2.5\Omega_{\s dr}$(a),
$\omega_{\s dr}=3.5\Omega_{\s dr}$ (b),
$\omega_{\s dr}=4.5\Omega_{\s dr}$ (c),
and $\omega_{\s dr}=5.6\Omega_{\s dr}$(d).  The peaks at $\omega=n\Omega_{\s dr}$ have gaussian shape. Their amplitudes
evolve with drive in {\em oscillating} fashion.     }
\label{4b}
\end{figure}

\section{Fast drive}
If the drive frequency is much bigger than the width  of distribution of the hyperfine fields,  there are
no realizations of ${\bm \Omega}_{\s N}$ which are in resonance with drive. For all realizations, the spin dynamics will be
affected by drive when the amplitude,  $\omega_{\s dr}$, becomes comparable to $\Omega_{\s dr}$. Suppose that
${\bm \omega}_{\s dr}$ is directed along $x$, while the projections of ${\bm \Omega}_{\s N}$ are equal to $\Omega_{\s N}^x$,
$\Omega_{\s N}^y$, and $\Omega_{\s N}^z$. Without drive, the two frequencies of the spin dynamics are $\omega=0$ and
$\omega =\Omega_{\s N}=
\Bigl[ \left(\Omega_{\s N}^x\right)^2+ \left(\Omega_{\s N}^y\right)^2+\left(\Omega_{\s N}^z\right)^2\Bigr]^{1/2}$.
The prime effect of a fast drive is that the frequency $\Omega_{\s N}$ transforms into
\begin{equation}
\label{lambda}
\lambda_{\s N}=
\Biggl[\left(\Omega_{\s N}^x\right)^2+ \left\{\left(\Omega_{\s N}^y\right)^2+\left(\Omega_{\s N}^z\right)^2\right\}
J_0^2\Bigl(\frac{\omega_{\s dr}}{\Omega_{\s dr}}\Bigr)\Biggr]^{1/2},
\end{equation}
where $J_0(x)$ is the zero-order Bessel function.
Besides, both frequencies $\omega=0$ and $\omega=\lambda_{\s N}$ acquire satellites at $\omega=n\Omega_{\s dr}$ and
$\omega=\lambda_{\s N}+n\Omega_{\s dr}$. The derivation of Eq. (\ref{lambda}) is given in the Appendix. According to
Eq. (\ref{lambda}), as the drive amplitude increases, $J_0\Bigl(\frac{\omega_{\s dr}}{\Omega_{\s dr}}\Bigr)$ falls off,
so that the spin precesses only in the $y$-$z$ plane.

The derivation of the noise spectrum in the case of a fast drive is in line with procedure employed in Sect II.
In the Appendix, along with deriving Eq. (\ref{lambda}), we find the solutions of the equations of motion of a driven spin
corresponding to the frequencies $\omega=0$ and $\omega=\lambda_{\s N}$ and present these solutions in the form of
combination of the normalized eigenvectors similar to Eq. (\ref{eigenvectors})
\begin{multline}
\label{eigenvectors2}
\begin{pmatrix} S_+\\ S_x\\ S_- \end{pmatrix} = B_+\begin{pmatrix} \mu_+ e^{i\lambda_{\s N} t +i\Phi(t)}
\\ \mu_x e^{i\lambda_{\s N} t} \\ \mu_- e^{i\lambda_{\s N} t -i\Phi(t)}\end{pmatrix}+
B_x\begin{pmatrix} \eta_+ e^{i\Phi(t)} \\ \eta_x  \\ \eta_- e^{-i\Phi(t)} \end{pmatrix}\\
 + B_-\begin{pmatrix} \nu_+ e^{-i\lambda_{\s N} t +i\Phi(t)} \\ \nu_x e^{-i\lambda_{\s N} t} \\
\nu_- e^{-i\lambda_{\s N} t -i\Phi(t)}\end{pmatrix},\quad\quad\quad\quad\quad
\end{multline}
where $S_{\pm}=\frac{1}{\sqrt{2}}(S_y\pm iS_z)$,
the oscillating phase $\Phi(t)$ is defined as
\begin{equation}
\label{phase}
\Phi(t)=\omega_{\s dr}\int_0^tdt'\cos(\Omega_{\s dr}t')=
\frac{\omega_{\s dr}}{\Omega_{\s dr}}\sin(\Omega_{\s dr}t),
\end{equation}
and the components of the eigenvectors are related as follows
\begin{equation}
\label{1}
\eta_x=\frac{\Omega_{\s N}^x}{\lambda_{\s N}},
\end{equation}
\begin{equation}
\label{2}
\mu_x=\nu_x=-i\eta_+=i\eta_-=
-\frac{\Bigl[(\Omega_{\s N}^y)^2+
(\Omega_{\s N}^z)^2\Bigr]^{\frac{1}{2}}}{\sqrt{2}\lambda_{\s N}}
 J_0\Bigl(\frac{\omega_{\s dr}}{\Omega_{\s dr}}\Bigr),
\end{equation}
\begin{equation}
\label{3}
\mu_+=-\nu_-=\frac{i}{2}\frac{\Bigl[(\Omega_{\s N}^y)^2+(\Omega_{\s N}^z)^2\Bigr]}
{\lambda_{\s N}(\lambda_{\s N}-\Omega_{\s N}^x)}J_0^2\Bigl(\frac{\omega_{\s dr}}{\Omega_{\s dr}}\Bigr),
\end{equation}
\begin{equation}
\label{4}
\mu_-=-\nu_+=\frac{i}{2}\frac{\Bigl[(\Omega_{\s N}^y)^2+(\Omega_{\s N}^z)^2\Bigr]}
{\lambda_{\s N}(\lambda_{\s N}+\Omega_{\s N}^x)}J_0^2\Bigl(\frac{\omega_{\s dr}}{\Omega_{\s dr}}\Bigr).
\end{equation}
The components $S_x$ of the eigenvectors are simple exponents. Then the contribution $\delta s_{x\omega}^2$ to the
noise spectrum follows directly from Eq. (\ref{eigenvectors2})
\begin{equation}
\label{x}
\delta s_{x\omega}^2=\frac{\pi}{2}
\Bigl[|\eta_x|^2\Delta(\omega)+|\mu_x|^2\Delta(\omega-\lambda_{\s N})+|\nu_x|^2\Delta(\omega+\lambda_{\s N})\Bigr].
\end{equation}
Concerning the contributions $\delta s_{y\omega}^2$ and $\delta s_{z\omega}^2$, they originate from the $S_+$ and $S_-$ components
of the eigenvectors, which are not simple exponents.
%Compared to Eq. (\ref{eigenvectors}) the elements of the eigenvectors Eq. (\ref{eigenvectors2}) are not the simple exponents.
This gives rise to the satellites spaced by $n\Omega_{\s dr}$ in the noise spectrum.
Relative  magnitudes of the satellites are found from the Fourier expansion
\begin{equation}
\label{expansion}
\exp(i\Phi(t))=\sum_n  J_n\Bigl(\frac{\omega_{\s dr}}{\Omega_{\s dr}}\Bigr)\exp(in\Omega_{\s dr}t).
\end{equation}
Since $\delta s_{y\omega}^2$ and $\delta s_{z\omega}^2$ give equal contributions to the ensemble-averaged spectrum,
it is convenient to average the combination $\delta s_{y\omega}^2+\delta s_{z\omega}^2$.
For this combination the result for a given hyperfine field assumes a compact form
\begin{multline}
\label{yz}
\delta s_{y\omega}^2+\delta s_{z\omega}^2=
\frac{\pi}{2} \Bigl[(|\eta_+|^2 + |\eta_-|^2)\sum_n{J_n^2\Bigl(\frac{\omega_{\s dr}}{\Omega_{\s dr}}\Bigr)\Delta(\omega-n\Omega_{\s dr})}
\\+(|\mu_+|^2 + |\mu_-|^2)\sum_n{J_n^2\Bigl(\frac{\omega_{\s dr}}{\Omega_{\s dr}}\Bigr)\Delta(\omega-\lambda_{\s N}-n\Omega_{\s dr})}+
\\(|\nu_+|^2 + |\nu_-|)^2\sum_n{J_n^2\Bigl(\frac{\omega_{\s dr}}{\Omega_{\s dr}}\Bigr)\Delta(\omega+\lambda_{\s N}+n\Omega_{\s dr})}\Bigr].
\end{multline}
From Eqs. (\ref{x}) and (\ref{yz}) we can trace the evolution of the averaged noise spectrum upon increasing the drive amplitude.
Firstly, in the weak-drive limit, $\omega_{\s dr} \ll \Omega_{\s dr}$, when the magnitudes of the satellites are negligible,
averaging of Eqs. (\ref{x}), (\ref{yz}) reproduces the result  Eq. (\ref{fromIvchenko}).
Indeed, when
$J_0\Bigl(\frac{\omega_{\s dr}}{\Omega_{\s dr}}\Bigr) \approx 1$,  the frequency $\lambda_{\s N}$ returns to $\Omega_{\s N}$.
The magnitudes, $|\eta_x|^2$ and $(|\eta_+|^2+|\eta_-|^2)$,  of a zero-frequency peaks in Eqs. (\ref{x}) and (\ref{yz})
become $(\Omega_{\s N}^x)^2/\Omega_{\s N}^2$ and  $[(\Omega_{\s N}^y)^2+(\Omega_{\s N}^z)^2]/\Omega_{\s N}^2$, as in the absence of drive.
Similarly, the fact that the magnitude of $\omega=\lambda_{\s N}$ peak assumes its zero-drive value follows from general relation
$|\mu_x|^2+|\mu_+|^2 + |\mu_-|^2=1$.

As the drive amplitude increases, the magnitude of a $\omega=0$ peak first decreases but eventually returns to its zero-drive value.
Indeed, in the limit $J_0\Bigl(\frac{\omega_{\s dr}}{\Omega_{\s dr}}\Bigr)\rightarrow 0$, we have $|\eta_x|^2 \approx 1$, while
$\eta_+$ and $\eta_-$ vanish. This suggests that the $\omega=n\Omega_{\s dr}$ satellites of a zero-frequency peak develop at
$\omega_{\s dr}\sim \Omega_{\s dr}$, but disappear in the strong-drive limit.  By contrast, the satellites at
$\omega=\pm\lambda_{\s N}+n\Omega_{\s dr}$ persist in the strong-drive limit. In this limit $\mu_x$ vanished, and thus we have
$|\mu_+|^2 + |\mu_-|^2\approx 1$. This suggests that all the noise power in $\omega=\lambda_{\s N}$ peak at
zero drive gets redistributed between the satellites at strong drive. Also, in the limit of strong drive, we have
$\lambda_{\s N}\approx |\Omega_{\s N}^x|$, so that Eq. (\ref{yz}) assumes the form
\begin{multline}
\label{last}
\delta s_{y\omega}^2+\delta s_{z\omega}^2=
\frac{1}{2}\sum_{n\neq 0}{J_n^2\Bigl(\frac{\omega_{\s dr}}{\Omega_{\s dr}}\Bigr)
\frac{\tau_s}{1+(\omega+\Omega_{\s N}^x+n\Omega_{\s dr})^2\tau_s^2}}.
\end{multline}
At zero drive the ensemble averaging
%of $\omega=\Omega_{\s N}$
over  $\Omega_{\s N}$
resulted in the noise spectrum given by $F(\omega)$, Eq. (\ref{distribution}).
By contrast, from Eq. (\ref{last}) we see that, for a strong drive, the ensemble averaging
of each term yields the distribution function of $\Omega_{\s N}^x$,
i.e. the shapes of the satellites are gaussian.
The overall noise spectrum in the presence of a fast drive is
illustrated in Figs. \ref{4a}, \ref{4b}.

\section{Discussion}

\begin{itemize}

\item
Our main result is Eq. (\ref{final}) for the averaged noise spectrum. This result
was obtained within the rotating wave approximation and applies for large enough
drive amplitudes $\omega_{\s dr}\tau_s \gg 1$.
Fig. \ref{Fig2} illustrates the evolution of the spectrum with $\omega_{\s dr}$.
As $\omega_{\s dr}$ increases, the magnitude of a central peak at $\omega=\Omega_{\s dr}$
grows linearly with drive, while the satellites at $|\omega \pm \Omega_{\s dr}|=\omega_{\s dr}$
broaden linearly with drive. Central peak and satellites merge at weak drive $\omega_{\s dr}\tau_s \sim 1$.
For smaller $\omega_{\s dr}$ the effect of drive on the spin dynamics is weak even for ``resonant" hyperfine
field configurations and can be treated perturbatively.  The effect of drive amounts to replacement
$\omega_{\s dr}$ by $\omega_{\s dr}^2\tau_s$ in the amplitude of the cental peak. The relative correction
to the background value of $\langle \delta s_{\omega}^2\rangle$ due to drive is
$\omega_{\s dr}^2\tau_s^2\ll 1$.  Despite being small, the effect of drive can be distinguished
in the derivative with respect to $\omega$. Indeed, the derivative of the background can be estimated as
$\frac{1}{\delta_e}F(\Omega_{\s dr})$, while the estimate for the derivative of the central peak is
$\omega_{\s dr}^2\tau_s^3F(\Omega_{\s dr})$. Thus the drive dominates the derivative in the domain
\begin{equation}
\omega_{\s dr}\tau_s \gg \frac{1}{\left(\delta_e\tau_s\right)^{1/2}}.
\end{equation}
Large typical value of the hyperfine field,  $\delta_e \gg \tau_s^{-1}$, which is presumed, allows to
distinguish the effect of drive even when it is weak.

\item
As in Refs. \onlinecite{Ivchenko}~-\onlinecite{Localized1},
we assumed that spin-relaxation time, $\tau_s$, resulting
from  random short-time correlated fields
%of the origin
different from hyperfine field, is the same for all elements
of the ensemble.
%$\delta$-correlated magnetic
%fields different from hyperfine field.
%We assumed that $\tau_s$ is the same
%for all elements of the ensemble.
ac-driven system is stationary but not
equilibrium. Still we calculated the noise
spectrum from eigenmodes. Justification for doing
this is that the temperature is much higher than all
the frequencies involved. Under this condition, all the
eigenmodes are equally represented in the spin
dynamics\cite{Localized}.

%is introduced with $\tau_s$ being the spin relaxation time of
%a single electron unrelated to the hyperfine or exchange
%interaction, considered as a phenomenological parameter
%of the theory and assumed to be the same for all
%the electrons. While deriving Eq. (5) it is assumed that
%the temperature of the system $T$ expressed in the energy
%units exceeds by far the characteristic interlevel energy
%splittings $|E_n - E_m|$. Hence, the averaging in Eq. (5)
%is performed with the equilibrium density matrix representing
%the equally occupied eigenstates. This condition
%is well satisfied in experiments on spin noise spectroscopy
%carried out at temperatures down to that of liquid Helium
%because the characteristic splittings between the n
%and m levels, due to both exchange and hyperfine interactions,
%correspond to $T = 10^{-3} K$.

\item
%The spin-relaxation rate, $\tau_s^{-1}$, being much smaller
%than the drive amplitude, $\omega_{\s dr}$, enters into the
%averaged noise spectrum only as a cutoff.
In a recent paper Ref. \onlinecite{ColoredNoise} a direct measurement of
the spin-relaxation rate, $\tau_s^{-1}$, was reported.
Such a measurement became possible due to implementing of the spin noise correlation
techniques, which involves two laser beams and allows to probe only specific configurations
of the hyperfine field. In this regard, the effect of the ac drive is prominent because it
also results from specific ``resonant" configurations.

\item
In  experiments on different semiconductor
structures\cite{Li1,NoiseInGaAs,NoiseInZnO,dot2010}
the measured width of the noise spectra ranged
from $\sim 2$MHz to $\sim 50$MHz.
Application of the ac drive with comparable
frequency does not constitute a problem, see e.g. Ref. \onlinecite{Boehme}.
It will require adding a coil to the conventional setup\cite{Li1,NoiseInGaAs,NoiseInZnO,dot2010}.

\item
Due to isotropy of the hyperfine fields the noise spectrum calculated above
does not depend on the direction of the ac magnetic field. This is the case
when the electron $g$-factor is isotropic. In experiment
Ref. \onlinecite{dot2010} it was established that the $g$-factor
is strongly anisotropic\cite{Remark}. This conclusion was drawn from the analysis of the shift
of the noise spectrum maximum with external magnetic field. With anisotropic $g$-factor,
drive-induced features of the noise spectrum will depend on the direction of $\omega_{\s dr}$.
While the position of a peak at $\omega=\Omega_{\s dr}$ is insensitive to the anisotropy,
the separation of the satellites will be bigger for the drive polarization along the bigger
$g$-value.

%The noise spectrum
%Despite the driving field  has a direction, ${\bm \omega}_{\s dr}$,
%the average noise spectrum is isotropic.
%External magnetic field,
%${\bm \Omega}$, breaks this isotropy even without drive.
%The contributions
%$\delta s_{x\omega}^2$,   $\delta s_{y\omega}^2$,
%and $\delta s_{z\omega}^2$, calculated assuming that the
%net magnetic field, ${\bm \Omega}+{\bm \Omega}_N$,
%is directed along the $z$-axis,
%enter into observable, mean square Faraday rotation,
%in combination
%\begin{equation}
%\label{fromIvchenko1}
%\frac{\tau_s}{2}\int d{\bm \Omega}_N F({\bm \Omega}_N)
%\Biggl[ \sin^2\theta(\delta s_{x\omega}^2+\delta s_{y\omega}^2)
%+\cos^2\theta \delta s_{z\omega}^2\Biggr],
%\end{equation}
%where $\theta$ is the angle between
%${\bm \Omega}+{\bm \Omega}_N$ and the direction of light propagation.
%Without magnetic field, the integration over $\theta$ reproduces
%Eq. (\ref{fromIvchenko}). For strong magnetic field, $\Omega \gg \delta_e$,
%Eq. (\ref{fromIvchenko1}) defines the ratio of magnitudes of the  peaks
%at $\omega=0$ and $\omega=\Omega$. In the latter case the effect of drive
%on the noise spectrum depends strongly on the relation between the drive frequency
%$\Omega_D$ and the Larmour frequency, $\Omega$, and on
% the mutual orientation of the
%driving field $2{\bm \omega_D}$ and the external field, $\Omega$.
%
%In the presence of external magnetic field
\end{itemize}

\section{Appendix}
Without loss of generality we can set $\Omega_{\s N}^y=0$.
We start from the equations of motion for the spin projections
%The  equations of
%motion for the spin projections in the presence of the linearly polarized drive,
%${\bf B}_1(t)={\bf i}B_{1}\cos(\omega t+\varphi)$, assume the form
\begin{align}
 \frac{\partial S_x}{\partial t} &=-\Omega_{\s N}^z S_y,\\
\frac{\partial S_y}{\partial t} &=-(\omega_{\s dr}\cos\Omega_{\s dr}t+\Omega_{\s N}^x) S_z+\Omega_{\s N}^z S_x,\\
\frac{\partial S_z}{\partial t} &=(\omega_{\s dr}\cos\Omega_{\s dr}t+\Omega_{\s N}^x) S_y.
\end{align}
To take advantage of the fact that the drive is fast
 it is convenient\cite{modulated} to switch to the new variables
\begin{align}
S_{x'} &= S_x, \\
S_{y'} &= S_y \cos \bigl(\Phi(t)+\Omega_{\s N}^x t\bigr) + S_z \sin \bigl(\Phi(t)+\Omega_{\s N}^x t\bigr) , \\
S_{z'} &= -S_y \sin \bigl(\Phi(t)+\Omega_{\s N}^x t\bigr) + S_z \cos \bigl(\Phi(t)+\Omega_{\s N}^x t\bigr),
\end{align}
where the phase $\Phi(t)$ is defined by Eq. (\ref{phase}).
%angle $\theta(t)$ is defined as
%\begin{equation}
%\theta(t)= \frac{\gamma B_1}{\omega} \sin (\omega t + \varphi)
%\end{equation}
The physical meaning of the above transformation is moving into the rotating frame
in which the ac field is canceled.
The equations of motion for the new variables read
\begin{align}
\label{InNewVariables}
\frac{\partial S_{x'}}{\partial t} &=\Omega_{\s N}^z S_{z'}\sin\bigl(\Phi(t)+\Omega_{\s N}^x t\bigr)-
\Omega_{\s N}^z S_{y'}\cos\bigl(\Phi(t)+\Omega_{\s N}^x t\bigr) , \nonumber \\
\frac{\partial S_{y'}}{\partial t} &=\Omega_{\s N}^z S_{x'}\cos\bigl(\Phi(t)+\Omega_{\s N}^x t\bigr), \nonumber \\
\frac{\partial S_{z'}}{\partial t} &=-\Omega_{\s N}^z S_{x'}\sin\bigl(\Phi(t)+\Omega_{\s N}^x t\bigr).
\end{align}
%One can see that there are two natural frequencies in the system Eq. (\ref{InNewVariables}), one is $\gamma B_0$ and the other is $\omeg
As a next step, we average Eqs. (\ref{InNewVariables})
over the time interval $\left(-\frac{\pi}{\Omega_{\s dr}}, \frac{\pi}{\Omega_{\s dr}}\right)$.
The justification for this step is that, since $\Omega_{\s dr} \gg \Omega_{\s N}$,
the spin projections do not change
significantly during this interval. Thus one can average only  $\cos\bigl(\Phi(t)+\Omega_{\s N}^x t\bigr)$
and $\sin\bigl(\Phi(t)+\Omega_{\s N}^x t\bigr)$
\begin{align}
\langle \cos\bigl(\Phi(t)+\Omega_{\s N}^x t\bigr) \rangle &=
J_0\Bigl(\frac{\omega_{\s dr}}{\Omega_{\s dr}}\Bigr) \cos\Omega_{\s N}^x t,\\
\langle \sin\bigl(\Phi(t)+\Omega_{\s N}^x t\bigr) \rangle &=
J_0\Bigl(\frac{\omega_{\s dr}}{\Omega_{\s dr}}\Bigr) \sin\Omega_{\s N}^x t.
\end{align}
It is also convenient to switch in the averaged equations to $S_{+'}=\frac{1}{\sqrt{2}}(S_{y'}+S_{z'})$ and
$S_{-'}=\frac{1}{\sqrt{2}}(S_{y'}-S_{z'})$. Then we get
%where $J_0(z)$ is a zero-order Bessel function,  we get
\begin{align}
\label{averaged}
\frac{\partial S_{x'}}{\partial t} &=
-\frac{1}{\sqrt{2}}\Omega_{\s N}^z J_0\Bigl(\frac{\omega_{\s dr}}{\Omega_{\s dr}}\Bigr)
\Bigl[S_{+'} e^{i\Omega_{\s N}^x t}+S_{-'} e^{-i\Omega_{\s N}^x t}\Bigr],\nonumber \\
\frac{\partial S_{+'}}{\partial t} &=
\frac{1}{\sqrt{2}}\Omega_{\s N}^z J_0\Bigl(\frac{\omega_{\s dr}}{\Omega_{\s dr}}\Bigr) S_{x'} e^{-i\Omega_{\s N}^x t},\nonumber \\
\frac{\partial S_{-'}}{\partial t} &=
\frac{1}{\sqrt{2}}\Omega_{\s N}^z J_0\Bigl(\frac{\omega_{\s dr}}{\Omega_{\s dr}}\Bigr) S_{x'} e^{i\Omega_{\s N}^x t}.
\end{align}
We see that the dynamics after averaging is slow, which justifies the averaging performed,
see Ref. \onlinecite{modulated} for rigorous justification. One can also see that Eqs. (\ref{averaged})
have the form of equations of motion in a constant magnetic field with $x$- and $z$-components being $\Omega_{\s N}^x$ and
$\Omega_{\s N}^zJ_0\Bigl(\frac{\omega_{\s dr}}{\Omega_{\s dr}}\Bigr)$, respectively. Finite $\Omega_{\s N}^y$ is naturally included as a $y$-component.
This immediately leads us to Eq. (\ref{lambda})
of the main text. Three eigenvectors correspond to rotations with frequencies $\omega=\lambda_{\s N}$, $\omega=0$,
and $\omega=-\lambda_{\s N}$.
To return to the lab frame one has to multiply $S_{+'}$ by $\exp\left(i\Phi(t)+i\Omega_{\s N}^xt\right)$
and $S_{-'}$ by $\exp\left(-i\Phi(t)-i\Omega_{\s N}^xt\right)$. This does not change the relation between the
components of the eigenvectors which have the form  Eq. (\ref{eigenvectors2}).

%Upon returning to the lab frame the solution of the system Eq. (\ref{averaged}) satisfying the condition $S_x(0)=1$ reads
%\begin{align}
%\label{labframe}
%S_x(t) &=  \cos\left[ \gamma B_0 J_0 \left( \frac{\gamma B_1}{\omega} \right) t \right], \\
%S_y(t) &= \sin\left[ \gamma B_0 J_0 \left( \frac{\gamma B_1}{\omega} \right) t \right]
%\cos \left[  \frac{\gamma B_1}{\omega} \sin(\omega t + \varphi) \right], \\
%S_z(t) &= \sin\left[ \gamma B_0 J_0 \left( \frac{\gamma B_1}{\omega} \right) t \right]
%\sin \left[  \frac{\gamma B_1}{\omega} \sin(\omega t + \varphi) \right].
%\end{align}
%As a final step, we average over the initial phase, $\varphi$, and arrive to Eq. (\ref{final}).

\section{Acknowledgements}
  We are grateful to C. Boehme, Y. Li, and E. G. Mishchenko for insightful discussions.
 This work was supported by NSF through MRSEC DMR-1121252.

\end{document}